\documentclass[aps,prb,reprint,superscriptaddress,longbibliography]{revtex4-2}
\usepackage{amsmath,amssymb,amsfonts,bm}
\usepackage{graphicx}
\usepackage{dcolumn}
\usepackage{siunitx}
\usepackage{color}
\usepackage{multirow}
\usepackage{hyperref}
\usepackage{tikz}

\graphicspath{{./figs/}}

\newcommand{\bfm}{\mathbf{m}}
\newcommand{\bfM}{\mathbf{M}}
\newcommand{\bfH}{\mathbf{H}}
\newcommand{\bfr}{\mathbf{r}}

\newcommand{\ms}{M_\mathrm{s}}
\newcommand{\LD}{L_\mathrm{D}}
\newcommand{\HD}{H_\mathrm{D}}
\newcommand{\ud}{\mathrm{d}}

\begin{document}

\title{Phase-Factor-Controlled Interaction and Bonding between a Chiral Bobber and a Skyrmion String in the Conical Phase}

\author{Haijun Zhao}
\thanks{Correspondence to: haijunzhao@seu.edu.cn}
\affiliation{Key Laboratory of Quantum Materials and Devices of Ministry of Education, School of Physics, Southeast University, Nanjing 211189, China}

\author{Tingting Yan}
\affiliation{Key Laboratory of Quantum Materials and Devices of Ministry of Education, School of Physics, Southeast University, Nanjing 211189, China}

\author{Shuai Dong}

\affiliation{Key Laboratory of Quantum Materials and Devices of Ministry of Education, School of Physics, Southeast University, Nanjing 211189, China}

\date{\today}

\begin{abstract}
Skyrmion interactions govern the formation of skyrmion lattices, clusters, and particle-like growth patterns.  In contrast, the interaction between a chiral bobber and a skyrmion string remains largely unexplored, despite the role of bobbers as intermediate states in skyrmion-string formation and annihilation.  Here we show that this interaction is intrinsically phase dependent in the conical phase.  Using three-dimensional micromagnetic simulations, we compute the locally relaxed constrained energy landscape as a function of the bobber--skyrmion separation $R$ and the surface phase factor $\phi_0$.  We find that changing $\phi_0$ qualitatively reshapes the interaction, producing repulsive, attractive, and bonding-like regimes that cannot be reduced to a conventional distance-dependent potential.  Real-space analysis shows that this behavior originates from phase-dependent reconstruction of nonaxisymmetric outer distortion shells.  The phase-controlled interaction persists over a finite field range and follows the expected top--bottom phase relation of surface-sensitive conical textures.  
These results identify the conical phase direction, often hidden in projected or thickness-averaged descriptions, as a previously underappreciated degree of freedom in the interaction between skyrmion strings and finite-length chiral textures, as demonstrated here for chiral bobbers.

\end{abstract}

\maketitle

\section{Introduction}

Magnetic skyrmions and related topological spin textures in chiral magnets have attracted sustained interest because of their nanometer-scale size, topological stability, and efficient current-driven dynamics~\cite{muhlbauer2009Skyrmion_,yu2010Realspace_,fert2013Skyrmions_,nagaosa2013Topological_,iwasaki2013Currentinduced_,jiang2017Direct_,zhao2025Controlling_,kim2025Topological_,kim2020Mechanisms_,kim2021Kinetics_}.  In magnets with bulk Dzyaloshinskii--Moriya interaction (DMI), the competition among exchange, DMI, Zeeman energy, and boundary-induced twists stabilizes a variety of three-dimensional (3D) magnetic objects.  These include skyrmion strings~\cite{rybakov2013Threedimensional_,brearton2022Threedimensional_,yu20243D_,jiang2024Thermal_,seki2022Direct_}, chiral bobbers~\cite{rybakov2015New_,zheng2018Experimental_,ran2021Creation_,redies2019Distinct_,seki2022Direct_,zhao2025Controlling_}, surface spirals~\cite{zhao2025Phasefactorcontrolled_}, stacked spirals~\cite{rybakov2016New_,turnbull2022Xray_}, and more complex hybrid or knotted topological strings~\cite{li2026Electrically_,seki2022Direct_}.  Such textures are of fundamental interest as 3D chiral solitons and may serve as building blocks for 3D spintronic architectures~\cite{fert2013Skyrmions_,zazvorka2019Thermal_,zhao2025Phasefactorcontrolled_,xia2023Universal_}.

Interactions between localized topological objects play a central role in their equilibrium arrangement and collective dynamics.  In effective particle descriptions, these interactions are commonly represented by pair potentials, which also provide microscopic input for particle-based or molecular-dynamics-type simulations~\cite{malescio2003Stripe_,zhao2012Analysis_,zhao2017Pattern_,edlund2010Universality_}.  For magnetic skyrmions, the interaction depends strongly on the background state.  In a saturated ferromagnetic background, skyrmions usually repel each other and form triangular lattices.  In the conical phase, the interaction becomes nonmonotonic: in our previous work, the repulsive-core/attractive-tail interaction between skyrmion strings was calculated and attributed to the reduction of the cone--skyrmion interfacial energy, explaining the particle-like growth of skyrmion crystals from the conical phase~\cite{kim2020Mechanisms_,kim2021Kinetics_}.  This picture is consistent with earlier theoretical and experimental studies of attractive skyrmion interactions in the conical phase~\cite{leonov2016Threedimensional_,du2018Interaction_} and with recent numerical studies of pairwise and many-body interactions~\cite{vizarim2026Manybody_}.  Even in the conical phase, however, the interaction between two extended skyrmion strings can still be largely described by an distance-dependent energy profile $E(R)$.

\begin{figure}[!t]
\centering
\includegraphics[width=0.98\linewidth]{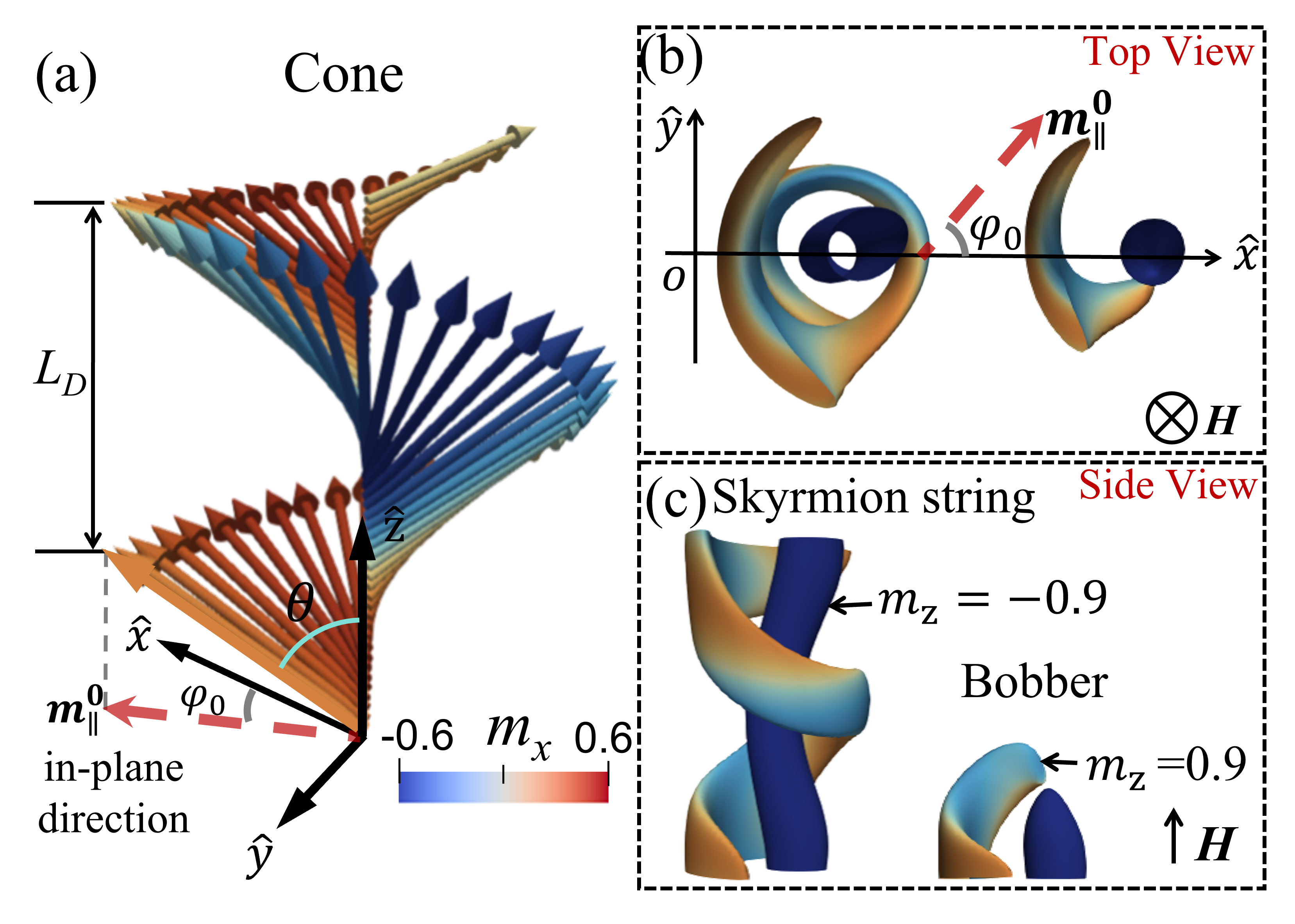}
\caption{
Basic spin textures in the conical background.
(a) Schematic illustration of the conical phase, whose surface phase factor $\phi_0$ selects a local in-plane reference direction.
(b,c) Top and side views of an isolated skyrmion string and a chiral bobber embedded in the conical phase.
The bobber has a length of approximately $\LD/2$ and is localized near the bottom surface.
The blue isosurfaces show $m_z=-0.9$ and visualize the down-magnetized cores of the skyrmion string and the bobber, while the colored isosurfaces show $m_z=0.9$.
The colors of the arrows in (a) and of the isosurfaces in (b,c) represent $m_x$, indicating the in-plane orientation of the local magnetization.
}
\label{fig1}
\end{figure}

Such a description becomes incomplete when the texture or the background supplies an additional in-plane direction.  Direction-dependent skyrmion interactions have been found, for example, when skyrmions are distorted by a tilted magnetic field or magnetocrystalline anisotropy~\cite{kameda2021Controllable_}.  The conical phase contains a more intrinsic source of directionality.  As sketched in Fig.~\ref{fig1}(a), its transverse magnetization rotates along the field direction, so the conical phase defines a local in-plane reference direction at a given surface.  Our previous thin-film calculations already showed a weak orientation dependence of the skyrmion--skyrmion interaction in the conical background~\cite{kim2020Mechanisms_}.  For extended skyrmion strings, this phase information is partly averaged along the film thickness or hidden in projected descriptions.  For finite-length textures, by contrast, such averaging is incomplete, and the local conical phase can become a direct variable of the interaction.

Chiral bobbers provide a concrete example of such finite-length textures.  In finite films, the DMI-induced boundary condition produces surface and edge twists~\cite{leonov2016Chiral_,hals2017New_,zhang2018Direct_,song2018Quantification_}, which support surface-localized textures such as surface spirals~\cite{zhao2025Phasefactorcontrolled_} and chiral bobbers~\cite{rybakov2015New_,zheng2018Experimental_,ran2021Creation_,redies2019Distinct_,seki2022Direct_,zhao2025Controlling_}.  As shown in Fig.~\ref{fig1}(b,c), a skyrmion string extends through the film thickness, whereas a chiral bobber is terminated by a Bloch point and remains localized near one surface.  This Bloch-point termination makes the bobber a natural intermediate state for changing the length and connectivity of a skyrmion string.  The motion of a Bloch point can elongate a bobber into a full skyrmion string~\cite{kim2020Mechanisms_}, while the nucleation and separation of Bloch points can break a skyrmion string into bobber-like segments during annihilation processes~\cite{zhao2025Controlling_}.  The bobber--skyrmion interaction is therefore directly relevant to skyrmion-string formation, skyrmion-crystal growth from the conical phase, and skyrmion annihilation in finite films~\cite{kim2020Mechanisms_,kim2021Kinetics_,zhao2025Controlling_}.

The role of the conical phase factor has recently been demonstrated in surface spirals, where the surface phase factor $\phi_0$ controls their appearance, penetration depth, bending direction, and single- or multiple-spiral configurations~\cite{zhao2025Phasefactorcontrolled_}.  Since a chiral bobber is also localized near the surface, its outer distortion shell can inherit the same local conical phase.  The interaction between a bobber and a skyrmion string should therefore depend not only on their separation $R$, but also on the relative orientation between the separation direction and the surface phase factor $\phi_0$.

Here we study this bobber--skyrmion interaction in the conical phase using 3D micromagnetic simulations.  By computing the locally relaxed constrained energy landscape as a function of $R$ and $\phi_0$, we show that the interaction cannot be reduced to a single distance-dependent pair potential.  Instead, the conical phase factor turns it into a phase-dependent energy landscape with distinct interaction regimes.  Real-space analysis shows that the strongest bonding-like transitions arise from phase-dependent matching and reconnection of nonaxisymmetric outer distortion shells.  These results identify the conical phase direction, usually hidden in projected or thickness-averaged descriptions, as an active degree of freedom in the interaction between skyrmion strings and finite-length chiral textures, as demonstrated here for chiral bobbers.

\section{Model and numerical methods}
\label{sec:model}

\subsection{Micromagnetic model and relaxation method}

We consider a bulk-DMI chiral magnet in a film geometry with free surfaces normal to the applied field.  Following earlier continuum micromagnetic studies of chiral surface twists, surface-localized spiral states, and conical-phase textures~\cite{leonov2016Chiral_,rybakov2016New_,zhao2025Controlling_,kim2025Topological_,zhao2025Phasefactorcontrolled_}, we use the minimal exchange--DMI--Zeeman energy functional
\begin{equation}
\varepsilon
=
A\sum_{\alpha=x,y,z}(\nabla m_\alpha)^2
+
D\,\bfm\cdot(\nabla\times\bfm)
-
\mu_0 \ms \bfH\cdot\bfm ,
\label{eq:energy_density}
\end{equation}
where $\bfm=\bfM/\ms$ is the unit magnetization, $A$ is the exchange stiffness, $D$ is the bulk-DMI constant, $\ms$ is the saturation magnetization, and $\bfH=H\hat{z}$ is the applied magnetic field.  We use the helical period $\LD=4\pi A/|D|$ and the saturation field $\HD$ as the characteristic length and field scales, and denote the reduced field by $h=H/\HD$.

Dipolar interactions are not included in the main calculations.  This choice allows us to isolate the minimal interaction mechanism arising from exchange, bulk DMI, Zeeman energy, and the chiral surface boundary condition.  In real films, magnetostatic effects can renormalize the cone angle, shift the stability range of short bobbers, modify the absolute energy scale, and affect the detailed shape of relaxed textures.  These effects may therefore lead to quantitative changes in the calculated energy landscapes.  However, the phase dependence studied here is mainly controlled by the relative orientation between the bobber--skyrmion pair and the transverse magnetization of the conical background.  This symmetry ingredient, together with the DMI-induced free-surface twist, is already contained in Eq.~\eqref{eq:energy_density}.  The calculated landscapes should therefore be viewed as a minimal micromagnetic description of the phase-controlled bobber--skyrmion interaction, rather than a material-specific prediction including all magnetostatic corrections.

The spin configurations are relaxed using the reduced Landau--Lifshitz--Gilbert dynamics,
\begin{equation}
\frac{\partial \bfm}{\partial \tau}
=
-\bfm\times\mathbf{h}_{\mathrm{eff}}
-
\alpha\,\bfm\times
\left(
\bfm\times\mathbf{h}_{\mathrm{eff}}
\right),
\label{eq:LLG}
\end{equation}
where $\mathbf{h}_{\mathrm{eff}}$ is the reduced effective field obtained from Eq.~\eqref{eq:energy_density}, and $\alpha$ is the damping constant.  In the present work, we are interested only in locally relaxed configurations rather than in real-time spin dynamics.  Therefore, in the actual relaxation process, the precessional first term in Eq.~\eqref{eq:LLG} is omitted and only the damping term is retained.  This damping-only procedure dissipates the micromagnetic energy and drives the system toward a locally stable configuration.  The resulting relaxed states are then used to evaluate the relative energy landscape of the constrained bobber--skyrmion pair.

\subsection{Geometry, conical phase, and phase convention}

The micromagnetic equations are solved using an in-house finite-difference micromagnetic code previously used in studies of helix reorientation~\cite{kim2025Topological_} and phase-factor-controlled surface states in chiral magnets~\cite{zhao2025Phasefactorcontrolled_}.
Unless otherwise stated, the simulation cell has an in-plane size of $L_x=6\LD$ and $L_y=3\LD$, and a thickness of $L_z=1.5\LD$.  The cell is discretized into $N_x\times N_y\times N_z=160\times80\times80$ grid points, corresponding to
\begin{equation}
\Delta x=\Delta y=\frac{3}{80}\LD,\qquad
\Delta z=\frac{3}{160}\LD .
\label{eq:mesh}
\end{equation}
The finer discretization along the $z$ direction is used to resolve the surface-localized bobber structure, the Bloch-point termination, and the associated surface distortion.  Periodic boundary conditions are imposed in the in-plane directions to approximate an extended film, while free boundary conditions are used at the top and bottom surfaces.  To check the robustness against discretization and lateral finite-size effects, we repeated selected calculations using a finer mesh and a larger in-plane simulation cell.  The main phase-sector classification and the occurrence of bonding/reconstructive transitions remain unchanged, indicating that the reported interaction landscapes are not controlled by the mesh size or by lateral finite-size effects.

For a given reduced field $h<1$, the reference background is the conical phase.  In the film geometry considered here, the system is translationally invariant in the $xy$ plane and has free surfaces normal to the $z$ direction.  The conical state with propagation vector along $z$ satisfies the natural boundary condition of the bulk-DMI model and can be written as
\begin{eqnarray}
\bfm_{\mathrm{c}}(z;\phi_0)
=
\sin\theta_{\mathrm{c}}
\left[
\cos\phi(z)\hat{x}
+
\sin\phi(z)\hat{y}
\right]
+
\cos\theta_{\mathrm{c}}\hat{z},
\label{eq:cone}
\end{eqnarray}
with
\begin{equation}
\phi(z)=\eta_D qz+\phi_0,
\qquad
q=\frac{2\pi}{\LD}.
\label{eq:cone_phase}
\end{equation}
Here $\theta_{\mathrm{c}}$ is the cone angle, $\cos\theta_{\mathrm{c}}=h$ in the reduced units used here, and $\eta_D$ specifies the handedness of the conical rotation for the sign convention in Eq.~\eqref{eq:energy_density}.  In the present calculations we take $D<0$, the same chirality convention as in Ref.~\cite{zhao2025Phasefactorcontrolled_}, and therefore $\eta_D=-1$.

The phase factor $\phi_0$ depends on the choice of the origin along the $z$ direction.  A shift of the coordinate origin changes $\phi_0$ by a constant, so $\phi_0$ itself is not an absolute observable.  For each surface considered below, we choose the coordinate origin such that the reference surface is located at $z=0$.  With this convention, $\phi_0$ denotes the in-plane direction of the transverse conical magnetization at that reference surface.  The physically relevant quantity is therefore the relative orientation between this surface phase direction and the ordered bobber--skyrmion separation direction.

The relative position of the two topological objects is fixed by convention.  We place the skyrmion string on the left and the bobber on the right, and define the ordered separation vector as
\begin{equation}
\mathbf{R}
=
\mathbf r_{\mathrm{bob}}
-
\mathbf r_{\mathrm{sk}}
=
R\hat{x}.
\label{eq:separation_vector}
\end{equation}
Thus the laboratory direction of the separation vector is fixed.  Scanning $\phi_0$ is equivalent to scanning the relative orientation between the bobber--skyrmion pair and the in-plane direction of the conical background at the reference surface.  If the laboratory direction of the separation vector is denoted by $\phi_R$, the relevant relative angle is $\phi_0-\phi_R$.  In the present convention, $\phi_R=0$ and is absorbed into the definition of $\phi_0$.  The relation to the opposite DMI chirality follows from a mirror transformation.  Reversing the sign of $D$ changes the handedness of the conical rotation and is equivalent to a mirror-related geometry in which the ordered separation direction is reversed, $\mathbf R\rightarrow-\mathbf R$.  Since the relevant angle is the relative orientation between the surface phase direction and the ordered separation vector, reversing $\mathbf R$ corresponds to a $180^\circ$ relabeling of the phase factor.  In the present convention, the opposite-chirality landscape is therefore obtained as
\begin{equation}
\Delta E_{D>0}(R,\phi_0)
=
\Delta E_{D<0}(R,\phi_0+180^\circ),
\label{eq:D_chirality_mapping}
\end{equation}
where the phase factor is understood modulo $360^\circ$.  Thus the opposite chirality does not generate an independent interaction landscape.

\subsection{Initial states and constrained interaction landscape}

A reference skyrmion string is generated in the conical background and then relaxed under the same field.  A chiral bobber is constructed by replacing a finite surface segment of the conical state with the corresponding segment of the relaxed skyrmion string, followed by full 3D relaxation.  This procedure produces a localized skyrmion-string fragment terminated by a Bloch point inside the sample and embedded in the same conical background.  Top-surface and bottom-surface bobbers are constructed in the same way, with the reference surface chosen accordingly.  We tested different initial bobber lengths and found that the shortest stable relaxed bobber has a length of about $\LD/2$; this shortest stable bobber is used throughout the main calculations unless otherwise stated.

A bobber--skyrmion pair state is constructed by placing the short bobber and the skyrmion string in the same conical background with a prescribed in-plane separation $R$.  Similar constrained-energy calculations, in which skyrmions are held at prescribed positions while the surrounding spin texture is relaxed, have been used to quantify skyrmion--skyrmion interaction potentials in two-dimensional chiral magnets under tilted fields or magnetocrystalline anisotropy~\cite{kameda2021Controllable_}.  Here, for each prescribed separation $R$ and conical phase factor $\phi_0$, the pair state is relaxed under a core-pinning constraint.  The core regions of the bobber and the skyrmion string are identified by the condition $m_z<-0.95$ and are used to define their positions.  During relaxation, the spins inside these small core regions are fixed to prevent translational motion driven by the mutual interaction, while the surrounding spin texture is allowed to relax freely.  The resulting state is therefore a locally relaxed constrained configuration at fixed $R$ and $\phi_0$.

The total energy of this constrained pair state is computed by integrating the micromagnetic energy density over the entire simulation cell,
\begin{equation}
E(R,\phi_0)
=
\int_V
\varepsilon[
\bfm(\bfr;R,\phi_0)
]
\,\ud V .
\label{eq:Epair}
\end{equation}
In the figures below, we plot the relative energy
\begin{equation}
\Delta E(R,\phi_0)
=
E(R,\phi_0)-E_0 ,
\label{eq:relative_energy}
\end{equation}
where $E_0$ is a constant offset chosen as the minimum value in the plotted data set unless otherwise stated.  This shift only fixes the energy zero and does not affect the shape of the interaction landscape.

The main interaction landscape is obtained by a constrained direct-relaxation method.  Each pair state with prescribed $R$ and $\phi_0$ is constructed independently and relaxed from its own initial configuration.  Therefore, each point in the energy landscape represents a locally relaxed constrained state obtained without using the relaxation history of neighboring points.

As shown below, in some phase sectors the direct-relaxation results exhibit abrupt bonding-like jumps, suggesting that more than one locally stable structural branch may coexist at the same separation.  To distinguish these branches and to examine the path dependence of the reconstruction, we additionally perform approaching and separating scans for selected values of $\phi_0$.  In these path-dependent scans, the relaxed configuration from the previous separation is used as the initial state for the next separation.

To realize such scans, the direct spin-fixing constraint on the bobber core is replaced by a localized auxiliary magnetic field applied parallel to the down-magnetized bobber core, i.e., opposite to the external field direction.  The auxiliary-field region is shifted step by step so that the bobber--skyrmion separation $R$ is decreased in the approaching scan or increased in the separating scan.  This procedure mimics a local probe that drags the bobber position while allowing the surrounding spin texture to relax.  The auxiliary field is used only to guide the bobber position during the path-dependent scans and is not included in the energy reported in the figures.  Comparing the direct-relaxation results with the approaching and separating scans allows us to identify coexisting locally relaxed branches and reconstructive transitions in the bonding sector.

We further checked representative bonding-sector cuts by varying the core-identification threshold and the size of the pinned core region.  These tests did not change the occurrence of the abrupt branch switching or the associated shell-reconstruction pattern, indicating that the bonding transition does not originate from a particular choice of the core mask or pinning constraint.

\begin{figure*}[htbp]
\centering
\includegraphics[width=0.98\linewidth]{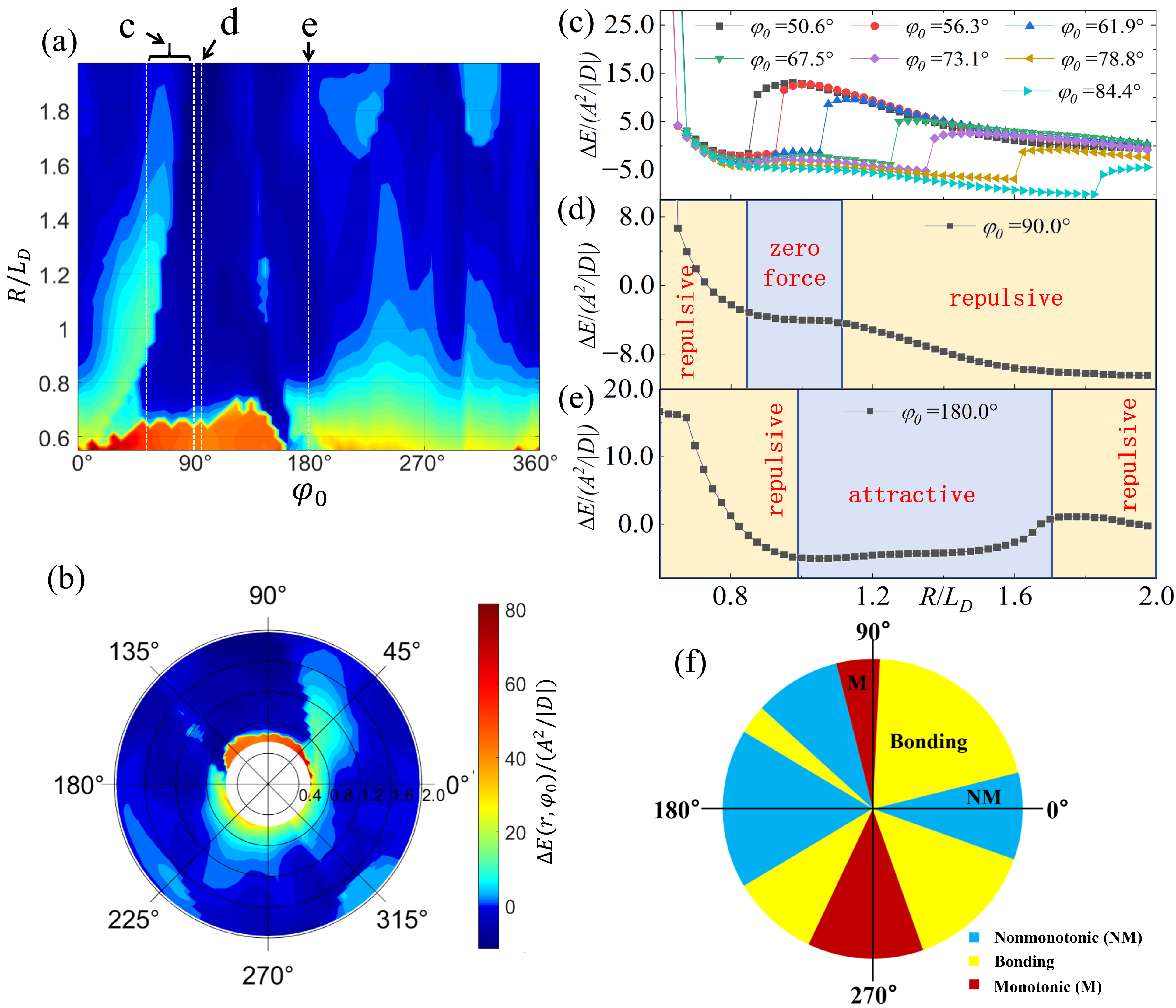}
\caption{
Phase-factor-dependent bobber--skyrmion interaction landscape.
(a) Contour map of the relative energy $\Delta E(R,\phi_0)$ in the separation--phase-factor plane.
Only the range $0.58\LD \le R \le 2\LD$ is shown in order to avoid the strongly repulsive short-distance region.
(b) Polar representation of the same energy landscape, where the radial coordinate is $R/\LD$ and the polar angle is the conical phase factor $\phi_0$.
(c--e) Representative one-dimensional energy cuts at fixed $\phi_0$, whose positions are marked in (a).
(c) Consecutive cuts for $\phi_0=50.7^\circ$--$84.4^\circ$, showing abrupt branch switching into lower-energy states.
The critical separation shifts to larger $R$ and the jump magnitude decreases as $\phi_0$ increases.
(d) Energy cut at $\phi_0=90^\circ$, showing a monotonic repulsive interaction with an extended nearly flat region where the effective radial force is small.
(e) Energy cut at $\phi_0=180^\circ$, showing a nonmonotonic interaction with a local minimum, corresponding to a repulsive--attractive--repulsive sequence.
(f) Classification of the fixed-$\phi_0$ energy cuts into monotonic continuous (M), nonmonotonic continuous (NM), and bonding/reconstructive branches.
All panels are plotted using the same relative-energy scale $\Delta E$.
}
\label{fig2}
\end{figure*}

\section{Results}
\label{sec:results}

\subsection{Orientation-dependent interaction landscape}

Figure~\ref{fig2} summarizes the bobber--skyrmion interaction landscape in the conical phase.  With the ordered separation direction fixed as defined above, varying the surface phase factor $\phi_0$ scans the relative orientation between the bobber--skyrmion pair and the in-plane direction of the conical background.  The interaction is therefore represented by the two-dimensional landscape $\Delta E(R,\phi_0)$.

Figure~\ref{fig2}(a) shows the contour map of $\Delta E(R,\phi_0)$ in the $(R,\phi_0)$ plane, and Fig.~\ref{fig2}(b) replots the same data in polar form.  The energy landscape is clearly not a simple distance-dependent pair potential.  Both the energy magnitude and the branch structure depend strongly on $\phi_0$, demonstrating that the conical phase factor acts as an orientational control parameter for the interaction.

The landscape contains three qualitatively different types of one-dimensional cuts at fixed $\phi_0$.  Some cuts are smooth and monotonic, some are smooth but nonmonotonic with a local minimum, and some display an abrupt drop into a lower-energy branch.  Representative cuts, whose positions are marked in Fig.~\ref{fig2}(a), are shown in Fig.~\ref{fig2}(c--e).
Figure~\ref{fig2}(c) shows consecutive cuts from the sector with abrupt energy drops, where $\phi_0$ varies from $50.7^\circ$ to $84.4^\circ$.  These jumps indicate a transition between two locally relaxed structural branches, and we therefore refer to this regime as the bonding or reconstructive sector.  The jump position and magnitude evolve systematically with $\phi_0$.  For smaller $\phi_0$, the transition occurs at shorter separation and is accompanied by a larger energy drop.  As $\phi_0$ increases, the transition shifts to larger $R$ and becomes weaker.  This continuous evolution indicates that the bonding behavior is not an isolated numerical event, but a finite angular sector controlled by the conical phase factor.
Figure~\ref{fig2}(d) shows a representative monotonic cut at $\phi_0=90^\circ$.  The energy remains monotonic over the scanned range and therefore corresponds to a repulsive interaction.  However, its slope becomes very small over an extended interval, giving a nearly force-free separation window within the monotonic regime.
Figure~\ref{fig2}(e) shows a representative nonmonotonic cut at $\phi_0=180^\circ$, where the energy develops a local minimum.  This corresponds to a repulsive--attractive--repulsive sequence as the bobber and skyrmion approach each other.

Based on these fixed-$\phi_0$ cuts, Fig.~\ref{fig2}(f) classifies the interaction into three regimes: monotonic continuous (M), nonmonotonic continuous (NM), and bonding/reconstructive branches.  
In the M regime, the energy varies smoothly and monotonically with separation and represents a repulsive interaction over the scanned range, although the slope can become very small for some phase factors.  
In the NM regime, the energy remains continuous but develops a local minimum.  Equivalently, the effective radial force, defined as $F_R=-\partial \Delta E/\partial R$, changes sign across the minimum.
In the bonding/reconstructive regime, the energy exhibits an abrupt branch switching associated with a structural reconstruction of the bobber--skyrmion pair.

\begin{figure*}[htbp]
\centering
\includegraphics[width=0.98\linewidth]{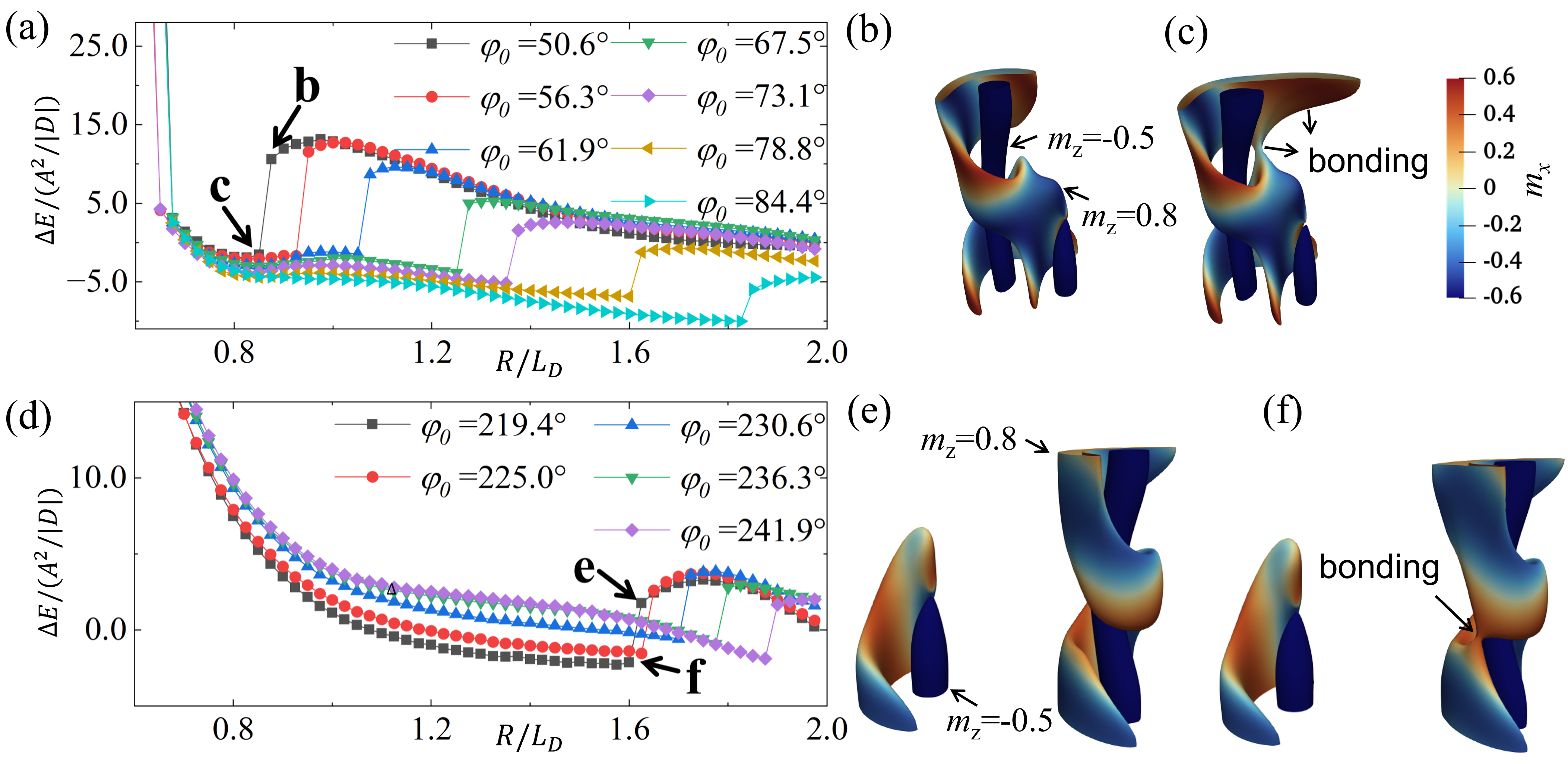}
\caption{
Bonding transition and real-space shell reconstruction.
(a) Representative bonding-sector energy cuts, with two configurations marked before and after an abrupt jump.
(b,c) Corresponding 3D spin configurations.
(d) Phase-shifted bonding sector with jumps around $R\simeq1.6\LD$.
(e,f) Configurations before and after the jump marked in (d).
The blue tubes show the $m_z=-0.5$ core isosurfaces, and the colored $m_z=0.8$ isosurfaces, colored by $m_x$, show the outer distortion shell.
In (b,c), the jump forms a bridge-like connection between the upper skyrmion-string shell and the lower bobber--string composite distortion.
In (e,f), the bridge-like connection forms mainly within the skyrmion-string outer shell near the bobber-termination region.
See Supplementary Movies~1 and 2 for the corresponding reconstruction processes.
}
\label{fig3}
\end{figure*}

\begin{figure}[!t]
\centering
\begin{tikzpicture}
\node[inner sep=0pt] (img) {\includegraphics[width=0.48\textwidth]{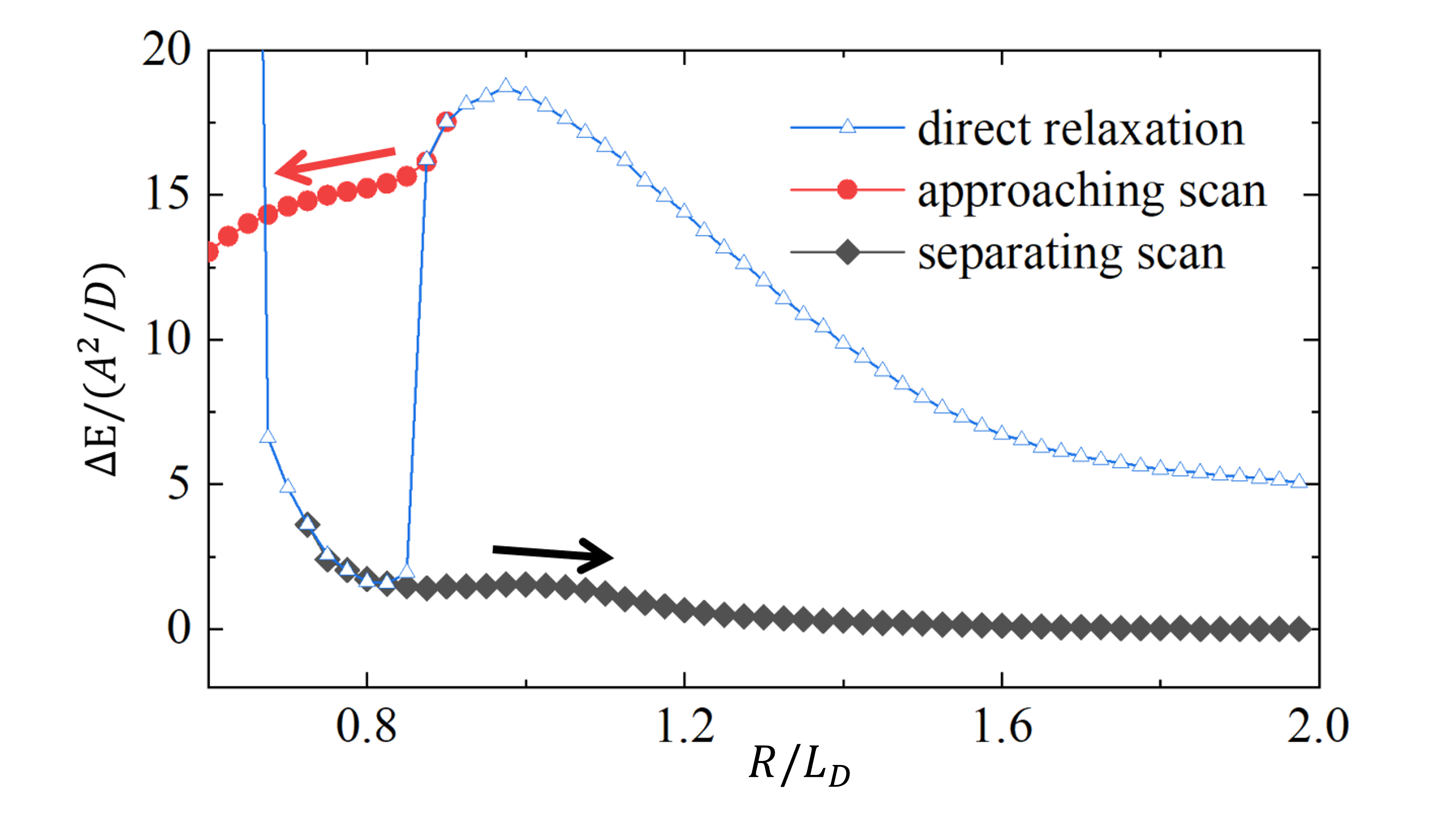}};
\node[
anchor=north west,
fill=white,
fill opacity=1,
text opacity=1,
inner sep=1.5pt
] at ([xshift=58mm,yshift=-29mm]img.north west)
{\footnotesize \shortstack[l]{$\phi_0=50.6^\circ$\\[-1pt]$H=0.45\HD$}};
\end{tikzpicture}
\caption{
Path-dependent relaxation across a representative bonding transition at $\phi_0=50.6^\circ$ and $H=0.45\HD$, corresponding to the cut highlighted in Fig.~\ref{fig3}(a).
The relative energy $\Delta E$ is shown for the direct-relaxation method, the approaching scan, and the separating scan, with arrows indicating the scan directions.
The different paths reveal hysteresis-like branch dependence and support the coexistence of unbonded and bonded locally relaxed states.
}
\label{fig4}
\end{figure}

Taken together, Fig.~\ref{fig2} shows that the bobber--skyrmion interaction in the conical phase is governed by the surface phase factor.  Changing $\phi_0$ not only changes the interaction strength, but also switches the qualitative branch type.  The interaction should therefore be viewed as a phase-factor-dependent landscape $\Delta E(R,\phi_0)$, rather than as a conventional distance-dependent pair potential.  In the following subsection, we show that the abrupt jumps in the bonding sector originate from real-space reconstruction of the surrounding distortion shell.

\subsection{Bonding transition and shell reconstruction}

We next examine the real-space origin of the abrupt energy jumps identified in the bonding sector.  Figure~\ref{fig3} compares two representative bonding processes.  Figure~\ref{fig3}(a) replots the bonding-sector energy cuts from Fig.~\ref{fig2}(c), with two configurations marked immediately before and after an abrupt jump.  The corresponding 3D spin textures are shown in Figs.~\ref{fig3}(b) and \ref{fig3}(c).  Figure~\ref{fig3}(d) shows another bonding sector shifted in phase, where the jumps occur at relatively large separations around $R\simeq1.6\LD$; the corresponding configurations are shown in Figs.~\ref{fig3}(e) and \ref{fig3}(f).  In all real-space plots, the blue tubes show the $m_z=-0.5$ isosurfaces and visualize the down-magnetized cores of the skyrmion string and the bobber, while the colored $m_z=0.8$ isosurfaces visualize the outer distortion shell, with color representing $m_x$.

The first bonding process is shown in Figs.~\ref{fig3}(b) and \ref{fig3}(c).  Before the jump, the lower part of the skyrmion-string outer shell already overlaps and merges with the bobber-induced outer shell, forming a local bobber--string composite distortion near the lower part of the sample.  However, because the bobber has a finite length, the upper part of the skyrmion-string shell has no corresponding bobber shell to merge with and remains as a separated spring-like segment.  After the jump, this upper spring-like segment becomes elongated and forms a bridge-like connection to the lower bobber--string composite distortion.  This new connection is absent in the isolated skyrmion string and lowers the energy of the pair state, giving rise to the bonded branch.  The same process is shown dynamically in Supplementary Movie~1.

The second bonding process, shown in Figs.~\ref{fig3}(e) and \ref{fig3}(f), has a different connection geometry.  In this phase sector, the bridge-like connection is formed mainly within the skyrmion-string outer shell itself, rather than between an upper string-shell segment and a lower bobber--string composite distortion.  The relevant shell segments are separated in the unbonded configuration, but become connected after the jump.  This connection again lowers the energy and produces a distinct bonded branch.  Compared with the first example, the reconnection occurs closer to the vertical level where the bobber terminates, near the Bloch-point region.  This reconstruction process is shown in Supplementary Movie~2.

These two examples show that the abrupt drop in $\Delta E(R,\phi_0)$ is caused by the formation of an energetically favorable bridge-like connection between outer-shell segments.  The bonding transition is not a conventional attraction between two rigid topological cores, but a reconstructive change of the surrounding distortion shell.  The bridge-like connection is controlled mainly by the phase-dependent outer distortion shells rather than by the nearly axisymmetric core regions.  Changing $\phi_0$ rotates the nonaxisymmetric shell distortions relative to the fixed bobber--skyrmion separation direction, so that different phase factors favor different shell-matching geometries and hence different bonded branches.

To further confirm the branch-switching interpretation, we perform path-dependent relaxation across a representative bonding transition.  Figure~\ref{fig4} compares the direct-relaxation result with approaching and separating scans for the cut at $\phi_0=50.6^\circ$ and $H=0.45\HD$.  In the approaching scan, the system follows the high-energy unbonded branch until this branch loses stability.  In contrast, the separating scan remains on the low-energy bonded branch over a broad range of separations.  This hysteresis-like behavior demonstrates the coexistence of two locally relaxed structural branches and confirms that the abrupt jump in the direct-relaxation curve corresponds to a transition between them.

\begin{figure*}[htbp]
  \centering
  \includegraphics[width=0.98\linewidth]{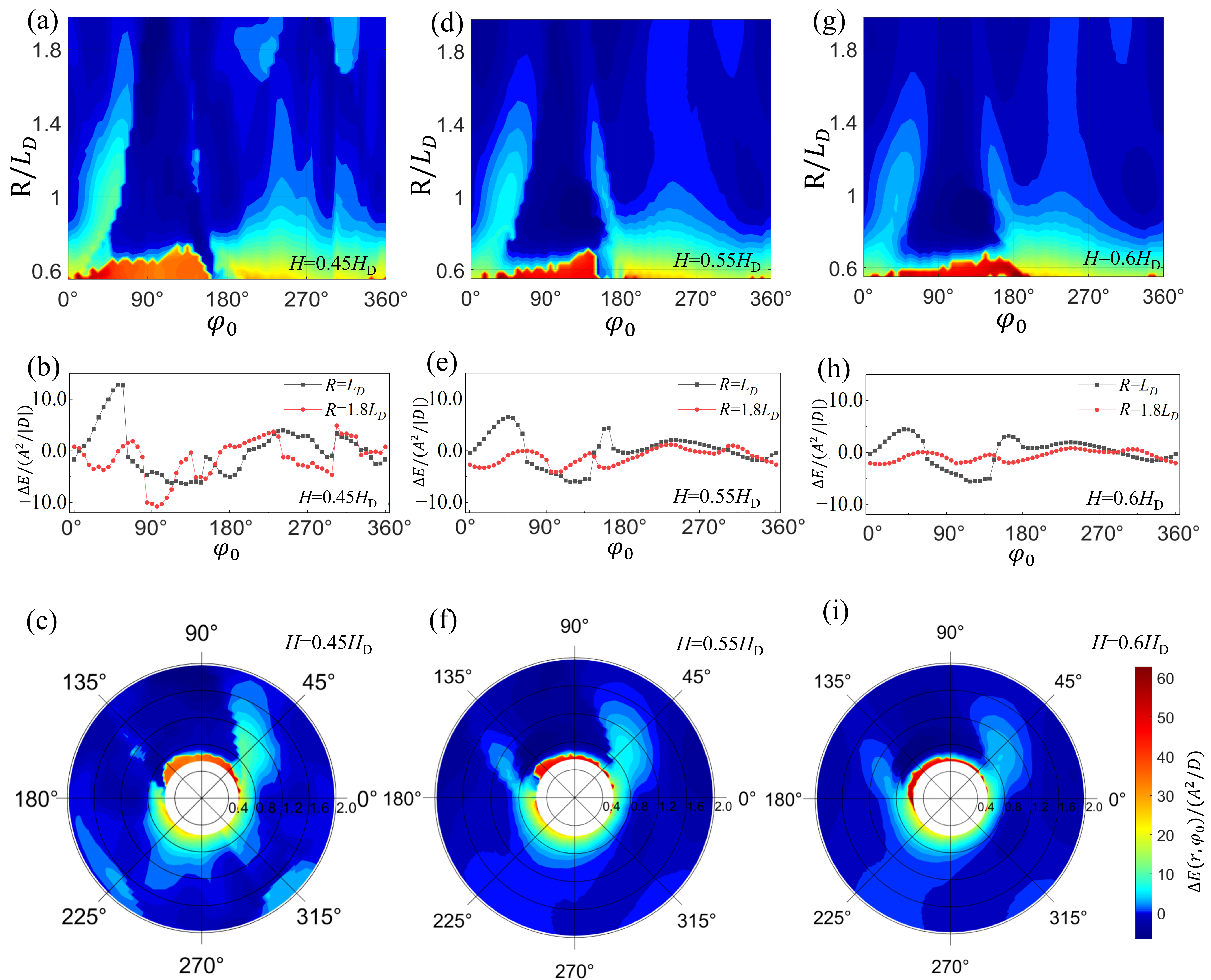}
\caption{
Field dependence of the phase-factor-dependent bobber--skyrmion interaction landscape.
The three columns correspond to $H=0.45\HD$, $0.55\HD$, and $0.60\HD$, respectively.
(a,d,g) Contour maps of the relative energy $\Delta E(R,\phi_0)$ in the separation--phase-factor plane.
(b,e,h) Angular energy profiles at two representative separations, $R=\LD$ and $R=1.8\LD$.
(c,f,i) Polar representations of the corresponding energy landscapes, where the radial coordinate is $R/\LD$ and the polar angle is $\phi_0$.
Increasing magnetic field suppresses the angular modulation and weakens the bonding sectors.
}
\label{fig5}
\end{figure*}

\begin{figure*}[htbp]
  \centering
  \includegraphics[width=0.98\linewidth]{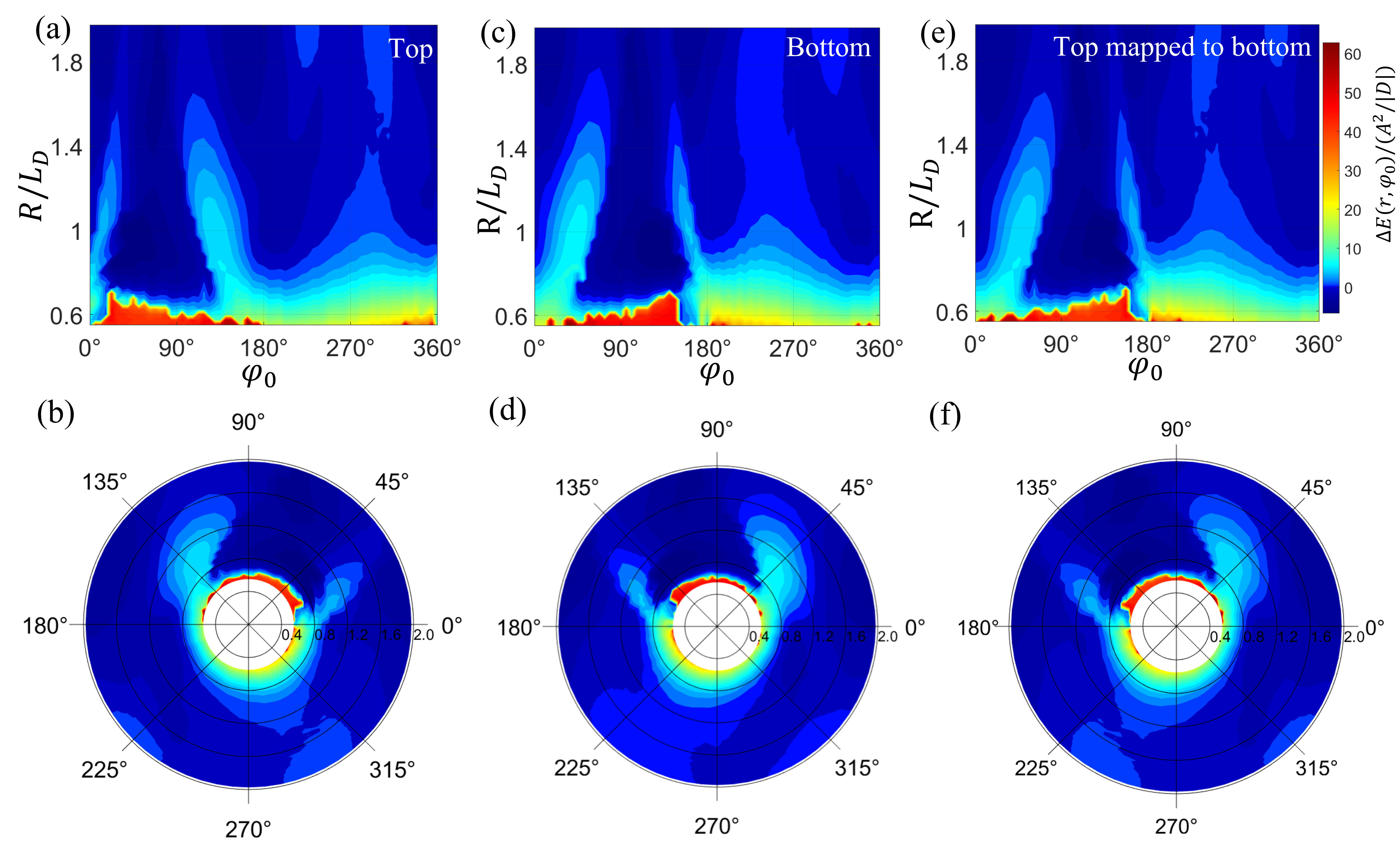}
\caption{
Top--bottom relation of the phase-factor-dependent bobber--skyrmion interaction at $H=0.55\HD$.
(a,b) Directly calculated bobber--skyrmion interaction landscape for a top-surface bobber, $\Delta E_{\mathrm{top}}(R,\phi_0)$, shown in Cartesian and polar representations.
(c,d) Directly calculated landscape for a bottom-surface bobber, $\Delta E_{\mathrm{bottom}}(R,\phi_0)$.
(e,f) Bottom-surface landscape predicted from the top-surface data by the phase mapping $\phi_0\rightarrow180^\circ-\phi_0$,
$
\Delta E_{\mathrm{top}\rightarrow\mathrm{bottom}}(R,\phi_0)
=
\Delta E_{\mathrm{top}}(R,180^\circ-\phi_0).
$
The close similarity between the mapped landscape in (e,f) and the directly calculated bottom-surface landscape in (c,d) demonstrates the top--bottom phase relation of the interaction.}
\label{fig6}
\end{figure*}

\subsection{Field dependence and top--bottom phase relation}
We next examine how the phase-factor-dependent interaction evolves with magnetic field.  Figure~\ref{fig5} compares the interaction landscapes at $H=0.45\HD$, $0.55\HD$, and $0.60\HD$.  The phase-dependent structure persists over the whole field range considered, but its angular modulation becomes progressively weaker as the field increases.  This trend is visible both in the two-dimensional maps and in the angular cuts at fixed separation.

The weakening of the angular modulation is consistent with the reduction of the cone angle at higher field.  Since the interaction depends on the compatibility between the bobber distortion, the skyrmion-string distortion, and the transverse component of the conical background, increasing the field reduces the phase-sensitive part of the interaction.  Consequently, the bonding and nonmonotonic sectors are most pronounced at $H=0.45\HD$ and become smoother at higher fields.  Nevertheless, the landscapes at different fields retain a similar angular organization, although with reduced modulation amplitude at larger fields.  This similarity indicates that the phase-factor-controlled interaction is a robust feature over the field range considered.

We finally examine how the bobber--skyrmion interaction changes when the bobber is attached to the opposite surface of the film.  Figure~\ref{fig6} compares the directly calculated landscapes for a skyrmion string interacting with a top-surface bobber and with a bottom-surface bobber at $H=0.55\HD$.  Motivated by the top--bottom phase relation of surface textures in the conical phase~\cite{zhao2025Phasefactorcontrolled_}, we test whether the two interaction landscapes are related by the same phase transformation,
\begin{equation}
\Delta E_{\mathrm{top}\rightarrow\mathrm{bottom}}(R,\phi_0)
=
\Delta E_{\mathrm{top}}(R,180^\circ-\phi_0).
\label{eq:top_bottom_mapping}
\end{equation}
 After this transformation, $\Delta E_{\mathrm{top}\rightarrow\mathrm{bottom}}(R,\phi_0)$ closely matches the directly calculated bottom-bobber landscape $\Delta E_{\mathrm{bottom}}(R,\phi_0)$.  The main angular features, including the bonding and nonmonotonic sectors, appear at nearly the same transformed phase factors.  Small quantitative differences remain, but they do not change the phase relation or the classification of the interaction regimes.  The overall correspondence demonstrates that reversing the bobber surface mainly changes the phase matching between the bobber distortion and the conical background.

The observed $180^\circ-\phi_0$ relation is therefore consistent with the surface phase relation found previously for phase-factor-controlled surface spirals~\cite{zhao2025Phasefactorcontrolled_}.  This agreement indicates that the angular anisotropy of the bobber--skyrmion interaction is not tied to a particular surface termination, but reflects the intrinsic phase-factor dependence of localized chiral textures embedded in the conical background.

\section{Conclusions}

We have studied the interaction between a short chiral bobber and a skyrmion string in the conical phase of a bulk-DMI chiral magnet.  Unlike the interaction between two extended skyrmion strings, which can still be largely described by an effective separation-dependent profile~\cite{kim2020Mechanisms_}, the bobber--skyrmion interaction depends strongly on the surface phase factor $\phi_0$ of the conical background.  The constrained energy landscape $\Delta E(R,\phi_0)$ therefore cannot be reduced to a conventional distance-dependent pair potential.  Depending on $\phi_0$, the interaction exhibits monotonic, nonmonotonic, and bonding/reconstructive regimes.  Real-space analysis shows that the abrupt bonding-like transitions arise from phase-dependent reconstruction of the nonaxisymmetric outer distortion shells surrounding the bobber--skyrmion pair.

This phase dependence persists over the magnetic-field range considered here.  Increasing the field reduces the cone angle and weakens the angular modulation, but the main phase-sector structure remains recognizable.  For bobbers attached to opposite film surfaces, the calculated landscapes are related by the transformation $\phi_0 \rightarrow 180^\circ-\phi_0$, consistent with the top--bottom phase relation previously found for phase-factor-controlled surface spirals.  This agreement shows that the angular anisotropy of the bobber--skyrmion interaction is not tied to a particular surface termination, but reflects the phase matching between the localized texture and the conical background.

These results identify the conical phase direction, usually hidden in projected or thickness-averaged descriptions, as an active degree of freedom in the interaction between skyrmion strings and finite-length chiral textures, as demonstrated here for chiral bobbers.  This perspective may be relevant to skyrmion-string formation and annihilation processes in which Bloch-point-terminated intermediate states appear near pre-existing skyrmions or skyrmion clusters.  More generally, it suggests that the phase of the conical background should be treated as a physical control parameter, rather than as a passive geometric label, when describing finite-length or surface-sensitive three-dimensional spin textures.

\section*{Data Availability}

The data that support the findings of this study are available from the corresponding author upon reasonable request.  The simulation code used in this work is not publicly available.

\begin{acknowledgments}
This work was supported by National Key Research 
and Development Program of China (No. 2025YFA1411100). Computer
resources at SEU were provided through Big Data Center
of Southeast University.

\end{acknowledgments}

\bibliographystyle{apsrev4-1}
\bibliography{all.bib}
\end{document}